\documentclass[sigconf]{acmart}

\usepackage{todonotes}
\usepackage{enumitem}
\usepackage{multirow}

\def\BibTeX{{\rm B\kern-.05em{\sc i\kern-.025em b}\kern-.08emT\kern-.1667em\lower.7ex\hbox{E}\kern-.125emX}}
    
\copyrightyear{2019}
\acmYear{2019}
\setcopyright{acmcopyright}
\acmConference[SIGIR '19]{Proceedings of the 42nd International ACM SIGIR Conference on Research and Development in Information Retrieval}{July 21--25, 2019}{Paris, France}
\acmBooktitle{Proceedings of the 42nd International ACM SIGIR Conference on Research and Development in Information Retrieval (SIGIR '19), July 21--25, 2019, Paris, France}
\acmPrice{15.00}
\acmDOI{10.1145/3331184.3331317}
\acmISBN{978-1-4503-6172-9/19/07}

\begin{document}

%
\title[CEDR: Contextualized Embeddings for Document Ranking]{CEDR: Contextualized Embeddings for Document Ranking}

%
\author{Sean MacAvaney}
\affiliation{\institution{IRLab, Georgetown University}}
\email{sean@ir.cs.georgetown.edu}

\author{Andrew Yates}
\affiliation{\institution{Max Planck Institute for Informatics}}
\email{ayates@mpi-inf.mpg.de}

\author{Arman Cohan}
\affiliation{\institution{Allen Institute for Artificial Intelligence}}
\email{armanc@allenai.org}

\author{Nazli Goharian}
\affiliation{\institution{IRLab, Georgetown University}}
\email{nazli@ir.cs.georgetown.edu}

%
\renewcommand{\shortauthors}{MacAvaney, et al.}

%
\begin{abstract}
Although considerable attention has been given to neural ranking architectures recently, far less attention has been paid to the term representations that are used as input to these models.
In this work, we investigate how two pretrained contextualized language models (ELMo and BERT) can be utilized for ad-hoc document ranking. Through experiments on \textsc{Trec} benchmarks, we find that several existing neural ranking architectures can benefit from the additional context provided by contextualized language models. Furthermore, we propose a joint approach that incorporates BERT's classification vector into existing neural models and show that it outperforms state-of-the-art ad-hoc ranking baselines. We call this joint approach CEDR (Contextualized Embeddings for Document Ranking). We also address practical challenges in using these models for ranking, including the maximum input length imposed by BERT and runtime performance impacts of contextualized language models.
\end{abstract}

%
\maketitle

\section{Introduction}
Recently, there has been much work designing ranking architectures to effectively score query-document pairs, with encouraging results~\cite{Guo2016ADR,Hui2018CoPACRRAC,Xiong2017EndtoEndNA}. Meanwhile, pretrained contextualized language models (such as ELMo~\cite{Peters:2018} and BERT~\cite{devlin2018bert}) have shown great promise on various natural language processing tasks~\cite{Peters:2018,devlin2018bert}. These models work by pre-training LSTM-based or transformer-based~\cite{Vaswani2017AttentionIA} language models on a large corpus, and then by performing minimal task fine-tuning (akin to ImageNet~\cite{deng2009imagenet,yosinski2014transferable}).

Prior work has suggested that contextual information can be valuable when ranking.
ConvKNRM~\cite{Dai2018ConvolutionalNN}, a recent neural ranking model, uses a convolutional neural network atop word representations, allowing the model to learn representations aware of context in \textit{local} proximity. In a similar vein, \citet{McDonald2018DeepRR} proposes an approach that learns a recurrent neural network for term representations, thus being able to capture context from the entire text~\cite{McDonald2018DeepRR}. These approaches are inherently limited by the variability found in the training data. Since obtaining massive amounts of high-quality relevance information can be difficult~\cite{Zamani2018SIGIR2W}, we hypothesize that \textit{pretrained} contextualized term representations will improve ad-hoc document ranking performance.

We propose incorporating contextualized language models into existing neural ranking architectures by using multiple similarity matrices -- one for each layer of the language model.
We find that, at the expense of computation costs, this improves ranking performance considerably, achieving state-of-the-art performance on the Robust 2004 and WebTrack 2012--2014 datasets.
We also show that combining each model with BERT's classification mechanism can further improve ranking performance.
We call this approach CEDR (Contextualzed Embeddings for Document Ranking).
Finally, we show that the computation costs of contextualized language models can be dampened by only partially computing the contextualized language model representations.
Although others have successfully used BERT for \textit{passage} ranking~\cite{Nogueira2019PassageRW} and question answering~\cite{2019EndtoEndOQ}, these approaches only make use of BERT's sentence classification mechanism. In contrast, we use BERT's term representations, and show that they can be effectively combined with existing neural ranking architectures.

In summary, our contributions are as follows:
\begin{enumerate}[noitemsep,topsep=0pt,leftmargin=*]
\item[-] We are the first to demonstrate that contextualized word representations can be successfully incorporated into existing neural architectures (PACRR~\cite{Hui2018CoPACRRAC}, KNRM~\cite{Xiong2017EndtoEndNA}, and DRMM~\cite{Guo2016ADR}), allowing them to leverage contextual information to improve ad-hoc document ranking.
\item[-] We present a new joint model that combines BERT's \textit{classification} vector with existing neural ranking architectures (using BERT's \textit{token} vectors) to get the benefits from both approaches.
\item[-] We demonstrate an approach for addressing the performance impact of computing contextualized language models by only partially computing the language model representations.
\item[-] Our code is available for replication and future work.\footnote{\url{https://github.com/Georgetown-IR-Lab/cedr}}
\end{enumerate}

\section{Methodology}

\subsection{Notation}

In ad-hoc ranking, documents are ranked for a given query according to a relevance estimate. Let $Q$ be a query consisting of query terms $\{q_1,q_2,...,q_{|Q|}\}$, and let $D$ be a document consisting of terms $\{d_1,d_2,...,d_{|D|}\}$. Let $ranker(Q,D)\in\mathbb{R}$ be a function that returns a real-valued relevance estimate for the document to the query. Neural relevance ranking architectures generally use a similarity matrix as input $\mathbf{S} \in \mathbb{R}^{|Q|\times|D|}$, where each cell represents a similarity score between the query and document: $\mathbf{S}_{i,j}=sim(q_i,d_j)$. These similarity values are usually the cosine similarity score between the word vectors of each term in the query and document.

\subsection{Contextualized similarity tensors}
Pretrained contextual language representations (such as those from ELMo~\cite{Peters:2018} and BERT~\cite{devlin2018bert}) are context sensitive; in contrast to more conventional pretrained word vectors (e.g., GloVe \cite{Pennington2014GloveGV}) that generate a single word representation for each word in the vocabulary, these models generate a representation of each word based on its context in the sentence. For example, the contextualized representation of word \textit{bank} would be different in \textit{bank deposit} and \textit{river bank}, while a pretrained word embedding model would always result in the same representation for this term. Given that these representations capture contextual information in the language, we investigate how these models can also benefit general neural ranking models.

Although contextualized language models vary in particular architectures, they typically consist of multiple stacked layers of representations (e.g., recurrent or transformer outputs). The intuition is that the deeper the layer, the more context is incorporated. To allow neural ranking models to learn which levels are most important, we choose to incorporate the output of all layers into the model, resulting in a three-dimensional similarity representation. Thus, we expand the similarity representation (conditioned on the query and document context) to $\mathbf{S_{Q,D}}\in\mathbb{R}^{L\times|Q|\times|D|}$ where $L$ is the number of layers in the model, akin to the channel in image processing. Let $context_{\mathbf{Q},\mathbf{D}}(t,l)\in\mathbb{R}^D$ be the contextualized representation for token $t$ in layer $l$, given the context of $Q$ and $D$.
Given these definitions, let the contextualized representation be:
\begin{equation}
\mathbf{S_{Q,D}}[l,q,d]=cos(context_{\mathbf{Q},\mathbf{D}}(q,l), context_{\mathbf{Q},\mathbf{D}}(d,l))
\end{equation}
for each query term $q\in Q$, document term $d\in D$, and layer $l\in [1..L]$. Note that when $q$ and $d$ are identical, they will likely not receive a similarity score of 1, as their representation depends on the surrounding context of the query and document. The layer dimension can be easily incorporated into existing neural models. For instance, soft n-gram based models, like PACRR, can perform convolutions with multiple input channels, and counting-based methods (like KNRM and DRMM) can count each channel individually.

\subsection{Joint BERT approach}

Unlike ELMo, the BERT model encodes multiple text segments simultaneously, allowing it to make judgments about text pairs. It accomplishes this by encoding two meta-tokens (\texttt{[SEP]} and \texttt{[CLS]}) and using text segment embeddings (\textit{Segment A} and \textit{Segment B}). The \texttt{[SEP]} token separates the tokens of each segment, and the \texttt{[CLS]} token is used for making judgments about the text pairs. During training, \texttt{[CLS]} is used for predicting whether two sentences are sequential -- that is, whether \textit{Segment A} immediately precedes \textit{Segment B} in the original text. The representation of this token can be fine-tuned for other tasks involving multiple text segments, including natural language entailment and question answering~\cite{2019EndtoEndOQ}.

We explore incorporating the \texttt{[CLS]} token's representation into existing neural ranking models as a joint approach. This allows neural rankers to benefit from deep semantic information from BERT in addition to individual contextualized token matches.

Incorporating the \texttt{[CLS]} token into existing ranking models is straightforward. First, the given ranking model produces relevance scores (e.g., n-gram or kernel scores) for each query term based on the similarity matrices. Then, for models using dense combination (e.g., PACRR, KNRM), we propose concatenating the \texttt{[CLS]} vector to the model's signals. For models that sum query term scores (e.g., DRMM), we include the \texttt{[CLS]} vector in the dense calculation of each term score (i.e., during combination of bin scores).

We hypothesize that this approach will work because the BERT classification mechanism and existing rankers have different strengths. The BERT classification benefits from deep semantic understanding based on next-sentence prediction, whereas ranking architectures traditionally assume query term repetition indicates higher relevance. In reality, both are likely important for relevance ranking.

\section{Experiment}

\subsection{Experimental setup}

\textbf{Datasets.}
We evaluate our approaches using two datasets: \textsc{Trec} Robust 2004 and WebTrack 2012--14. For Robust, we use the five folds from~\cite{huston2014parameters} with three folds used for training, one fold for testing, and the previous fold for validation. For WebTrack, we test on 2012--14, training each year individually on all remaining years (including 2009--10), and validating on 2011. (For instance, when testing on WebTrack 2014, we train on 2009--10 and 2012--13, and validate on 2011.) Robust uses \textsc{Trec} discs 4 and 5\footnote{520k documents; \url{https://trec.nist.gov/data_disks.html}}, WebTrack 2009--12 use ClueWeb09b\footnote{50M web pages, \url{https://lemurproject.org/clueweb09/}}, and WebTrack 2013--14 uses ClueWeb12\footnote{733M web pages, \url{https://lemurproject.org/clueweb12/}} as document collections.
We evaluate the results using the nDCG@20 / P@20 metrics for Robust04 and nDCG@20 / ERR@20 for WebTrack.

\textbf{Models.} Rather than building new models, in this work we use existing model architectures to test the effectiveness of various input representations. We evaluate our methods on three neural relevance matching methods: PACRR~\cite{Hui2018CoPACRRAC}, KNRM~\cite{Xiong2017EndtoEndNA}, and DRMM~\cite{Guo2016ADR}. Relevance matching models have generally shown to be more effective than semantic matching models, while not requiring massive amounts of behavioral data (e.g., query logs). For PACRR, we increase $k_{max}=30$ to allow for more term matches and better back-propagation to the  language model.

\textbf{Contextualized language models.}
We use the pretrained ELMo (Original, 5.5B) and BERT (BERT-Base, Uncased) language models in our experiments. For ELMo, the query and document are encoded separately. Since BERT enables encoding multiple texts at the same time using \textit{Segment~A} and \textit{Segment~B} embeddings, we encode the query (\textit{Segment~A}) and document (\textit{Segment~B}) simultaneously. Because the pretrained BERT model is limited to 512 tokens, longer documents are split such that document segments are split as evenly as possible, while not exceeding the limit when combined with the query and control tokens. (Note that the query is always included in full.) BERT allows for simple classification fine-tuning, so we also experiment with a variant that is first fine-tuned on the same data using the Vanilla BERT classifier (see baseline below), and further fine-tuned when training the ranker itself.

\textbf{Training and optimization.} We train all models using pairwise hinge loss~\cite{Dehghani2017NeuralRM}. Positive and negative training documents are selected from the query relevance judgments (positive documents limited to only those meeting the re-ranking cutoff threshold $k$ using BM25, others considered negative).
We train each model for 100 epochs, each with 32 batches of 16 training pairs. Gradient accumulation is employed when the batch size of 16 is too large to fit on a single GPU. We re-rank to top $k$ BM25 results for validation, and use P@20 on Robust and nDCG@20 on WebTrack to select the best-performing model. We different re-ranking functions and thresholds at test time for each dataset: BM25 with $k=150$ for Robust04, and QL with $k=100$ for WebTrack. The re-ranking setting is a better evaluation setting than ranking all qrels, as demonstrated by major search engines using a pipeline approach~\cite{Rosset2018OptimizingQE}. All models are trained using Adam~\cite{Kingma2015AdamAM} with a learning rate of 0.001 while BERT layers are trained at a rate of 2e-5.\footnote{Pilot experiments showed that a learning rate of 2e-5 was more effective on this task than the other recommended values of 5e-5 and 3e-5 by~\cite{devlin2018bert}.} Following prior work~\cite{Hui2018CoPACRRAC}, documents are truncated to 800 tokens.

\begin{table*}
\centering\small
\caption{\small
Ranking performance on Robust04 and WebTrack 2012--14. Significant improvements to [B]M25, [C]onvKNRM, [V]anilla BERT, the model trained with [G]lOve embeddings, and the corresponding [N]on-CEDR system are indicated in brackets (paired t-test, $p<0.05$).
}
\label{tab:results_main}
\vspace{-0.5em}
\scalebox{0.85}{
\begin{tabular}{llrrrr}
\toprule
& & \multicolumn{2}{c}{Robust04} & \multicolumn{2}{c}{WebTrack 2012--14} \\
 \cmidrule(lr){3-4} \cmidrule(lr){5-6}
Ranker & Input Representation & P@20 & nDCG@20 & nDCG@20 & ERR@20 \\
\midrule

BM25 & n/a & 0.3123 & 0.4140 & 0.1970 & 0.1472 \\
SDM~\cite{Metzler2005AMR} & n/a & 0.3749 & 0.4353 & - & - \\
\texttt{TREC-Best} & n/a & \bf0.4386 & \bf0.5030 & 0.2855 & \bf0.2530 \\
ConvKNRM & GloVe & 0.3349 & 0.3806 & [B] 0.2547 & [B] 0.1833 \\
Vanilla BERT & BERT (fine-tuned) & [BC] 0.4042 & [BC] 0.4541 & \bf[BC] 0.2895 & [BC] 0.2218 \\
\midrule
PACRR & GloVe & 0.3535 & [C] 0.4043 & 0.2101 & 0.1608 \\
PACRR & ELMo & [C] 0.3554 & [C] 0.4101 & [BG] 0.2324 & [BG] 0.1885 \\
PACRR & BERT & [C] 0.3650 & [C] 0.4200 & 0.2225 & 0.1817 \\
PACRR & BERT (fine-tuned) & [BCVG] 0.4492 & [BCVG] 0.5135 & [BCG] 0.3080 & [BCG] 0.2334 \\
CEDR-PACRR & BERT (fine-tuned) & \bf[BCVG] 0.4559 & \bf[BCVG] 0.5150 & \bf[BCVGN] 0.3373 & \bf[BCVGN] 0.2656 \\
\midrule
KNRM & GloVe & 0.3408 & 0.3871 & [B] 0.2448 & 0.1755 \\
KNRM & ELMo & [C] 0.3517 & [CG] 0.4089 & 0.2227 & 0.1689 \\
KNRM & BERT & [BCG] 0.3817 & [CG] 0.4318 & [B] 0.2525 & [B] 0.1944 \\
KNRM & BERT (fine-tuned) & [BCG] 0.4221 & [BCVG] 0.4858 & [BCVG] 0.3287 & [BCVG] 0.2557 \\
CEDR-KNRM & BERT (fine-tuned) & \bf[BCVGN] 0.4667 & \bf[BCVGN] 0.5381 & \bf[BCVG] 0.3469 & \bf[BCVG] 0.2772 \\
\midrule
DRMM & GloVe & 0.2892 & 0.3040 & 0.2215 & 0.1603 \\
DRMM & ELMo & 0.2867 & 0.3137 & [B] 0.2271 & 0.1762 \\
DRMM & BERT & 0.2878 & 0.3194 & [BG] 0.2459 & [BG] 0.1977 \\
DRMM & BERT (fine-tuned) & [CG] 0.3641 & [CG] 0.4135 & [BG] 0.2598 & [B] 0.1856 \\
CEDR-DRMM & BERT (fine-tuned) & \bf[BCVGN] 0.4587 & \bf[BCVGN] 0.5259 & \bf[BCVGN] 0.3497 & \bf[BCVGN] 0.2621 \\

\bottomrule
\end{tabular}
}
\end{table*}

\textbf{Baselines.}
We compare contextualized language model performance to the following strong baselines:
\begin{enumerate}[noitemsep,topsep=0pt,leftmargin=*]
\item[-] BM25 and SDM~\cite{Metzler2005AMR}, as implemented by Anserini~\cite{Yang2017AnseriniET}. Fine-tuning is conducted on the test set, representing the maximum performance of the model when using static parameters over each dataset.\footnote{$k_1$ in 0.1--4 (by 0.1), $b$ in 0.1--1 (by 0.1), SDM weights in 0--1 (by 0.05).} We do not report SDM performance on WebTrack due to its high cost of retrieval on the large ClueWeb collections.
\item[-] Vanilla BERT ranker. We fine-tune a pretrained BERT model (\texttt{BERT-Base, Uncased}) with a linear combination layer stacked atop the classifier \texttt{[CLS]} token. This network is optimized the same way our models are, using pairwise cross-entropy loss and the Adam optimizer. We use the approach described above to handle documents longer than the capacity of the network, and average the \texttt{[CLS]} vectors from each split.
\item[-] \texttt{TREC-best}: We also compare to the top-performing topic TREC run for each track in terms of nDCG@20. We use \texttt{uogTrA44xu} for WT12 (\cite{Limsopatham2012UniversityOG}, a learning-to-rank based run), \texttt{clustmrfaf} for WT13 (\cite{Raiber2013TheTA}, clustering-based), \texttt{UDInfoWebAX} for WT14 (\cite{Liu2014EntityCT}, entity expansion), and \texttt{pircRB04t3} for Robust04 (\cite{Kwok2004TREC2R}, query expansion using Google search results).\footnote{We acknowledge limitations of the TREC experimentation environment.}
\item[-] ConvKNRM~\cite{Dai2018ConvolutionalNN}, our implementation with the same training pipeline as the evaluation models.
\item[-] Each evaluation model when using GloVe~\cite{Pennington2014GloveGV} vectors.\footnote{\texttt{glove.42B.300d}, \url{https://nlp.stanford.edu/projects/glove/}}

\end{enumerate}

\subsection{Results \& analysis}

Table~\ref{tab:results_main} shows the ranking performance using our approach. We first note that the Vanilla BERT method significantly outperforms the tuned BM25 [V] and ConvKNRM [C] baselines on its own. This is encouraging, and shows the ranking power of the Vanilla BERT model. When using contextualized language term representations without tuning, PACRR and DRMM performance is comparable to that of GloVe [G], while KNRM sees a modest boost. This might be due to KNRM's ability to train its matching kernels, tuning to specific similarity ranges produced by the models. (In contrast, DRMM uses fixed buckets, and PACRR uses maximum convolutional filter strength, both of which are less adaptable to new similarity score ranges.) When fine-tuning BERT, all three models see a significant boost in performance compared to the GloVe-trained version. PACRR and KNRM see comparable or higher performance than the Vanilla BERT model. This indicates that fine-tuning contextualized language models for ranking is important. This boost is further enhanced when using the CEDR (joint) approach, with the CEDR models always outperforming Vanilla BERT [V], and nearly always significantly outperforming the non-CEDR versions [N]. This suggests that term counting methods (such as KNRM and DRMM) are complementary to BERT's classification mechanism. Similar trends for both Robust04 and WebTrack 2012--14 indicate that our approach is generally applicable to ad-hoc document retrieval tasks.

To gain a better understanding of how the contextual language model helps enhance the input representation, we plot example similarity matrices based on GloVe word embeddings, ELMo representations (layer 2), and fine-tuned BERT representations (layer 5). In these examples, two senses of the word \textit{curb} (restrain, and edge of street) are encountered. The first is relevant to the query (it's discussing attempts to restrain population growth). The second is not (it discusses street construction). Both the ELMo and BERT models give a higher similarity score to the correct sense of the term for the query.  This ability to distinguish different senses of terms is a strength of contextualized language models, and likely can explain some of the performance gains of the non-joint models.

\begin{figure}
\centering
\includegraphics[scale=0.63]{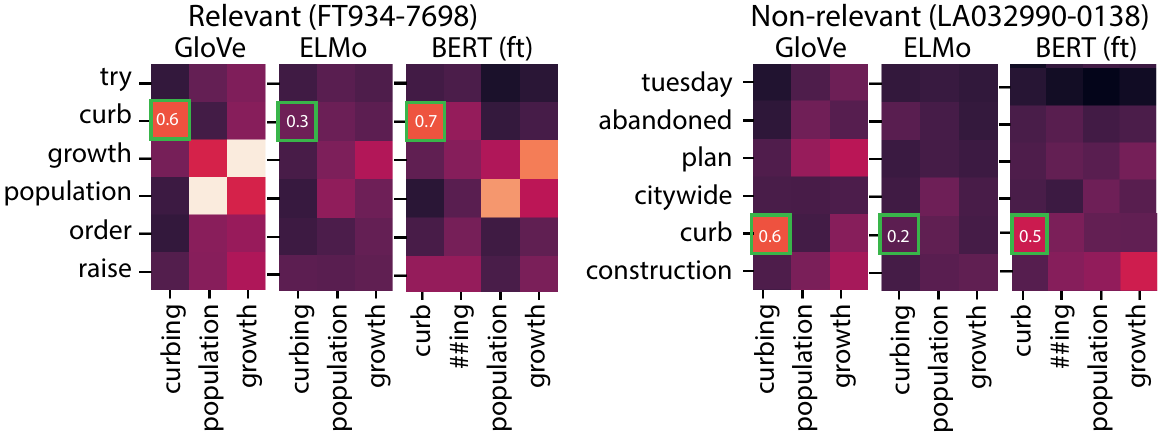}
\vspace{-1em}
\caption{Example similarity matrix excerpts from GloVe, ELMo, and BERT for relevant and non-relevant document for Robust query 435. Lighter values have higher similarity.}
\label{fig:simmat-ex}
\vspace{-0.9em}
\end{figure}

\begin{figure}
\includegraphics[scale=0.32]{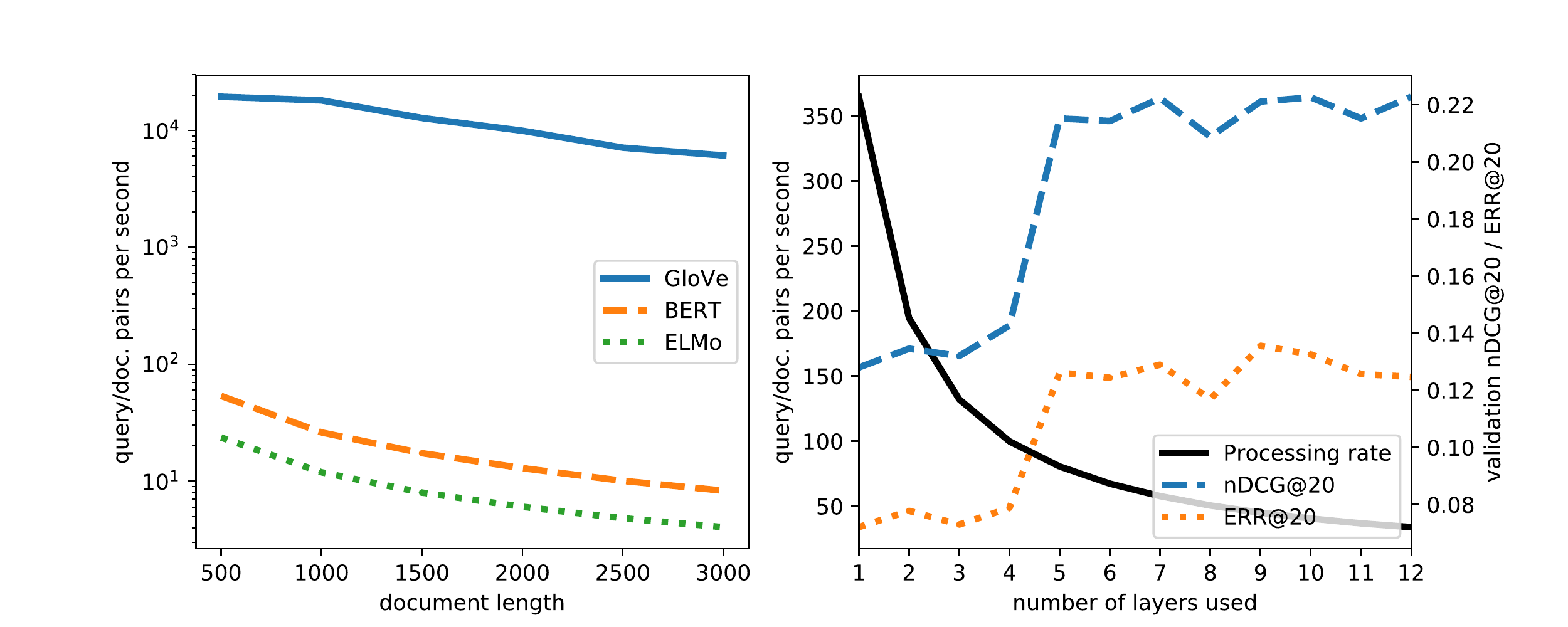}

\vspace{-0.4em}
\small(a)\hspace{12em}(b)
\vspace{-0.9em}

\caption{(a) Processing rates by document length for GloVe, ELMo, and BERT using PACRR. (b) Processing rate and dev performance of PACRR when using a subset of BERT layers.}
\label{fig:runtimes}
\vspace{-1.8em}
\end{figure}

Although the contextualized language models yield ranking performance improvements, they come with a considerable cost at inference time---a practical issue ignored in previous ranking work~\cite{Yang2017AnseriniET,Nogueira2019PassageRW}. To demonstrate this, in Figure~\ref{fig:runtimes}(a) we plot the processing rate of GloVe, ELMo, and BERT.\footnote{Running time measured on single GeForce GTX 1080 Ti GPU, data in memory.} Note that the processing rate when using static GloVe vectors is orders of magnitude faster than when using the contextualized representations, with BERT outperforming ELMo because it uses the more efficient Transformer instead of an RNN. In an attempt to improve the running time of these systems, we propose limiting the number of layers processed by the model. The reasoning behind this approach is that the lower the layer, the more abstract the matching becomes, perhaps becoming less useful for ranking. We show the runtime and ranking performance of PACRR when only processing only up to a given layer in Figure~\ref{fig:runtimes}(b). It shows that most of the performance benefits can be achieved by only running BERT through layer 5; the performance is comparable to running the full BERT model, while running more than twice as fast. While we acknowledge that our research code is not completely optimized, we argue that this approach is generally applicable because the processing of these layers are sequential, query-dependent, and dominate the processing time of the entire model. This approach is a simple time-saving measure.

\section{Conclusion}

We demonstrated that contextualized word embeddings can be effectively incorporated into existing neural ranking architectures and suggested an approach for improving runtime performance by limiting the number of layers processed.

%
\bibliographystyle{ACM-Reference-Format}
\bibliography{biblio}

\end{document}